\documentclass[french]{article}
\usepackage[T1]{fontenc}
\usepackage[utf8]{inputenc}
\usepackage{lmodern}
\usepackage[a4paper]{geometry}
\usepackage{babel}
\usepackage{datetime}
\usepackage{fancyhdr}

\usepackage{amssymb}
\usepackage{stmaryrd}
\usepackage{amsmath}
\usepackage{amsthm}
\usepackage{mathtools}
\usepackage{dsfont}
\usepackage{hyperref}
\usepackage{cleveref}
\usepackage{cancel}
\usepackage{graphicx}
\usepackage[svgnames]{xcolor}
\usepackage{proof}
\usepackage{wrapfig}

\usepackage{tikz}
\usepackage{tikz-cd}
\usepackage{tikzit} 
\usepackage{tikz-qtree}

\usepackage[backend=biber]{biblatex}
\theoremstyle{plain}
\newtheorem{thm}{Théorème}[section]
\newtheorem{lem}[thm]{Lemme}
\newtheorem{prop}[thm]{Proposition}

\theoremstyle{definition}

\newtheorem{defn}[thm]{Définition}
\newtheorem{exmp}[thm]{Exemple}

\theoremstyle{remark}
\newtheorem*{nb}{N.B}

\newenvironment{dem}[1][]{\paragraph*{\textnormal{\textit{Démonstration#1.}}}}{\hfill\qed\newline}

\usepackage{contour}
\usepackage[normalem]{ulem}

\contourlength{1pt}

\newcommand{\C}{\mathcal{C}}

\newcommand{\D}{\mathcal{D}}

\newcommand{\M}{\mathcal{M}}

\renewcommand{\O}{\mathcal{O}}

\renewcommand{\P}{\mathcal{P}}
 
\renewcommand{\S}{\mathcal{S}}

\newcommand{\V}{\mathcal{V}}

\newcommand{\X}{\mathcal{X}}

\newcommand{\sep}{\mid}

\newcommand{\name}{Pierre Lafourcade}
\author{\name}

\begin{document}

\begin{center}
	{\Huge Parties optimales dans Room 25 en modes solo et coopération} \\[1ex]
	{\Large Pierre Lafourcade}\\ [2ex]
	{\textit{Univ. Bordeaux, CNRS, Bordeaux INP, LaBRI, UMR 5800, F-33400 Talence, France}}
\end{center}

\begin{abstract}
	Nous étudions la question des parties optimales pour les modes solo et coopération du jeu Room 25 (saison 1). Nous démontrerons qu’il est impossible de remporter la partie en un seul tour, quelque soit la configuration de départ, mais qu’en revanche il est possible de le faire en deux pour certaines configurations. Nous proposerons une ouverture qui permet de gagner en deux tours si l’on a suffisamment de chance, tout en n’ayant qu’une faible probabilité de mener à une défaite immédiate. Nous verrons ensuite que l’on peut remporter la partie en un seul tour si l’on change très légèrement les règles, quoique la probabilité de succès est alors nettement plus faible que celle de la stratégie en deux tours. Enfin, nous prouverons que si au contraire les joueurs sont maximalement malchanceux, ils perdront quelle que soit leur stratégie.
\end{abstract}

\section{Introduction}

Room 25 est un jeu de société créé en 2013 par François Rouzé et édité par Matagot \autocite{Room25,Room25Saison}. Il met en scène un jeu télévisé dans lequel les candidats essaient de s’échapper d’un complexe mortel dont la topographie change au fur et à mesure de la partie. Nous nous focalisons ici sur la version originale, également appelée \textit{saison 1}.\\

Il existe plusieurs modes de jeu mais nous n’étudierons que les modes solo et coopération, qui ne diffèrent quasiment que par leur nombre de joueurs. Ces modes rendent le jeu moralement à un joueur puisque, même s’il y en a plusieurs, ils coopèrent et n’ont pas d’informations cachée des autres.\footnote{En principe le partage d’informations entre joueurs est partiel mais cela ne nous importe pas ici, comme nous verrons en \cref{ssec_actions}.} Nous étudierons des parties optimales, à savoir: combien de tours au minimum faut-il pour remporter la partie? Précisons que, le jeu étant à la fois soumis au hasard et à une information incomplète, les résultats présentés ici ne permettent pas de remporter une partie typique.\\

A notre connaissance, aucun travail théorique n’a encore été menés sur ce jeu.\\

Nous présenterons en \cref{sec_notations} les règles élémentaires du jeu ainsi qu’une notation algébrique. En \cref{sec_opti} nous présenterons et démontrerons les résultats d’optimalité à proprement parler. La \cref{sec_proba} étudiera les probabilités de succès des stratégies données dans la section précédente, ainsi que celles de défaite immédiate. Enfin en \cref{sec_imposs}, nous étudierons les cas de figure où, dans un certain sens que nous définirons, le hasard est optimalement défavorable.

\section{Notation algébrique}\label{sec_notations}

Cette section a pour objet d’introduire une notation algébrique qui servira ensuite à décrire des parties. Nous présenterons au passages les règles nécessaires aux preuves, en invitant le lecteur intéressé à se référer vers le manuel du jeu pour les règles complètes.

\subsection{Plateau}

Le plateau du jeu est composé de 25 tuiles carrées disposées en une grille $5\times 5$. Toutes les tuiles sont amenées à bouger selon un mécanisme qui sera expliqué plus tard, à l’exception de la tuile de départ, notée $\D$, qui ne peut en aucun cas quitter sa position centrale. Etant donnés l’importance de cette tuile à position fixe et la symétrie centrale du plateau, nous optons pour un repère cartésien dont l’origine est au centre du plateau. La tuile $\D$ a donc toujours la coordonnée $[0;\!0]$. Les lignes et les colonnes sont désignées par $[;y]$ et $[x;]$, $x$ ou $y$ étant la valeur commune à toutes les coordonnées de la ligne ou colonne en question.\\

Les tuiles représentent des salles dans lesquelles évoluent des personnages. Elles sont de différents types, que nous pouvons résumer ici par leurs couleurs, qui représentent globalement leurs effets:
\begin{itemize}
	\item les tuiles \textit{vertes} ont un effet positif ou neutre pour les personnages qui s’y trouvent;
	\item les tuiles \textit{jaunes} ont un effet négatif mais me peuvent en aucun cas provoquer la mort d’un personnage;
	\item les tuiles \textit{rouges} toutes un effet qui peut conduire à la mort d’un occupant, de manière plus ou moins conditionnelle;
	\item les tuiles \textit{bleues} sont la tuile de départ $\D$ où se situent tous les personnages au début de la partie, et la tuile de sortie $\S$, qu’ils doivent atteindre pour la remporter.\footnote{Cette dernière salle s’appelle \textit{Room 25} dans les règles du jeu et lui donne son nom, cependant nous l’appelons \textit{salle de sortie} ici pour rappeler sa fonction et pouvoir la désigner par l’abréviation $\S$.}
\end{itemize}
Le plateau doit impérativement contenir les deux tuiles bleues mais les 23 autres sont choisies lors de la mise en place parmi un ensemble plus grand. Le livret de règles donne une composition recommandée pour chaque mode de jeu (dont une commune aux modes solo et coopération), mais invite les joueurs à varier le jeu en choisissant eux-mêmes les 23 tuiles non-bleues du complexe.\\

\begin{wrapfigure}{r}{0.45\linewidth}
	\vspace{-10mm}
	\centering
	\begin{tikzpicture}[scale=.5]
		\begin{pgfonlayer}{nodelayer}
			\node [style=none] (0) at (-5, 5) {};
			\node [style=none] (1) at (5, -5) {};
			\node [style=none] (2) at (-1, 1) {};
			\node [style=none] (3) at (1, -1) {};
			\node [style=none] (4) at (-5.5, 3.25) {};
			\node [style=none] (5) at (-5.5, 0.75) {};
			\node [style=none] (6) at (5.25, 0.75) {};
			\node [style=none] (7) at (5.25, 3.25) {};
			\node [style=none] (8) at (-5.25, 5.25) {};
			\node [style=none] (9) at (-5.25, -5.25) {};
			\node [style=none] (10) at (-2.75, -5.25) {};
			\node [style=none] (11) at (-2.75, 5.25) {};
			\node [style=none] (12) at (0, 0) {$\D$};
			\node [style=none] (13) at (-5.5, 2) {};
			\draw [color=red] (13) node[left] {$[;\!1]$};
			\node [style=none] (13b) at (5.5, 2) {};
			\draw [color=red] (13b) node[right] {\phantom{$[;\!1]$}};
			\node [style=none] (14) at (-4, 5.25) {};
			\draw [color=Green] (14) node[above] {$[-2;]$};
			\node [style=none] (15) at (-3, 5) {};
			\node [style=none] (16) at (-1, 5) {};
			\node [style=none] (17) at (1, 5) {};
			\node [style=none] (18) at (3, 5) {};
			\node [style=none] (19) at (5, 5) {};
			\node [style=none] (20) at (-5, 3) {};
			\node [style=none] (21) at (-3, 3) {};
			\node [style=none] (22) at (-1, 3) {};
			\node [style=none] (23) at (1, 3) {};
			\node [style=none] (24) at (3, 3) {};
			\node [style=none] (25) at (5, 3) {};
			\node [style=none] (26) at (-4, 2) {$[-2;\!1]$};
			\node [style=none] (27) at (-5, 1) {};
			\node [style=none] (28) at (-3, 1) {};
			\node [style=none] (29) at (1, 1) {};
			\node [style=none] (30) at (3, 1) {};
			\node [style=none] (31) at (5, 1) {};
			\node [style=none] (32) at (5, -1) {};
			\node [style=none] (33) at (3, -1) {};
			\node [style=none] (34) at (-1, -1) {};
			\node [style=none] (35) at (-3, -1) {};
			\node [style=none] (36) at (-5, -1) {};
			\node [style=none] (37) at (-5, -3) {};
			\node [style=none] (38) at (-3, -3) {};
			\node [style=none] (39) at (-1, -3) {};
			\node [style=none] (40) at (1, -3) {};
			\node [style=none] (41) at (3, -3) {};
			\node [style=none] (42) at (5, -3) {};
			\node [style=none] (43) at (3, -5) {};
			\node [style=none] (44) at (1, -5) {};
			\node [style=none] (45) at (-1, -5) {};
			\node [style=none] (46) at (-3, -5) {};
			\node [style=none] (47) at (-5, -5) {};
		\end{pgfonlayer}
		\begin{pgfonlayer}{edgelayer}
			\fill [color=gray] (0.center) rectangle (1.center);
			\fill [color=cyan] (2.center) rectangle (3.center);
			\fill [color=LightGrey] (16.center) -- (0.center) -- (27.center) -- (28.center) -- (21.center) -- (22.center) -- cycle;
			\fill [color=LightGrey] (19.center) -- (17.center) -- (23.center) -- (24.center) -- (30.center) -- (31.center) -- cycle;
			\fill [color=LightGrey] (32.center) -- (33.center) -- (41.center) -- (40.center) -- (44.center) -- (1.center) -- cycle;
			\fill [color=LightGrey] (35.center) -- (36.center) -- (47.center) -- (45.center) -- (39.center) -- (38.center) -- cycle;
			\draw (0.center) to (19.center);
			\draw (20.center) to (25.center);
			\draw (31.center) to (27.center);
			\draw (36.center) to (32.center);
			\draw (42.center) to (37.center);
			\draw (47.center) to (1.center);
			\draw (1.center) to (19.center);
			\draw (18.center) to (43.center);
			\draw (44.center) to (17.center);
			\draw (16.center) to (45.center);
			\draw (46.center) to (15.center);
			\draw (0.center) to (47.center);
			\draw [color=red](7.center) rectangle (5.center);
			\draw [color=Green](8.center) rectangle (10.center);
		\end{pgfonlayer}
	\end{tikzpicture}
	\caption{Plateau initial avec la tuile $\D$ face recto et les autres face verso. Les cases claires sont les positions possible de $\S$.}
	\label{fig_plateau}
\end{wrapfigure}
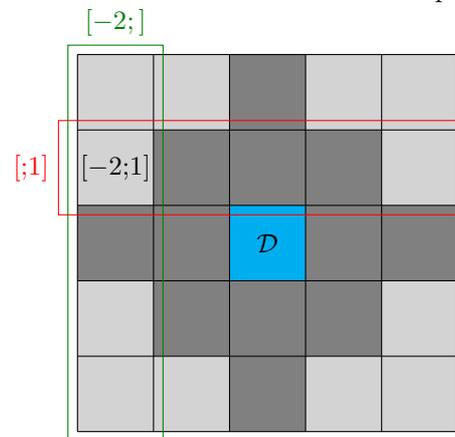

Nous utiliserons par la suite deux tuiles vertes particulières: la salle vide, notée $\V$, qui n’a aucun effet; et la salle de contrôle, notée $\C$ dont l’effet sera détaillé dans le paragraphe \textit{contrôler} de la \cref{ssec_actions}.\\

Au début de la partie, toutes les tuiles à l’exception de $\D$ sont face cachée, et la tuile $\S$ ne peut se trouver que dans une des douze positions près des coins, en gris clair sur la \cref{fig_plateau}. L’un des ressorts principaux du jeu est précisément le fait d’acquérir des informations sur les tuiles face verso pour atteindre $\S$ tout en évitant les salles dangereuses. Cependant nous considérons ici des parties optimales, ce qui nous fait supposer une disposition de plateau idéale, et \textit{a fortiori} nous permet d’ignorer la prise d’informations. Nous étudierons cependant dans un second temps la probabilité de réussite des stratégies, ainsi que celle d’aboutir à la mort d’au moins un personnage en essayant de la mettre à exécution dans une partie réelle (donc aléatoire).

\subsection{Personnages et actions}\label{ssec_actions}

Chaque joueur contrôle un ou plusieurs personnages (selon le nombre de joueurs) tout au long de la partie. Tous les personnages sont sur la tuile $\D$ au début de la partie. A chaque tour, les joueurs programment une ou deux actions distinctes pour chaque personnage. Les quatre actions possibles sont détaillées ci-après; précisons que deux tuiles sont dites \textit{adjacentes} ssi elles partagent une arête: en particulier deux cases en diagonale ne sont pas adjacentes.

\paragraph{Regarder (\textit{R})}\phantom{}\\
Le joueur regarde en secret le recto d’une tuile face verso adjacente au personnage effectuant l’action, puis la remet en place. La règle stipule que le joueur a interdiction de \og communiquer le nom de la salle \fg{}, mais qu’il peut \og donner une indication sur la dangerosité de la tuile \fg{}. Nous supposerons que des partenaires habiles sauront donner une telle indication qui soit suffisamment précise pour que les autres joueurs sachent, au sein de la stratégie commune, précisément comment aborder la tuile. Cette interprétation est sujette à débat, certes, mais elle est simplificatrice en ce qu’elle nous permet de traiter le mode coopération comme étant formellement à un seul joueur, et nous importe peu puisque nous considérons des parties théoriques dans lesquelles l’on peut se permettre de se fier à la chance pour ne pas avoir à si soucier de prendre des informations.

Cette action est désignée par la lettre $R$; le fait que le personnage $1$ regarde la tuile de coordonnées $[1;\!2]$ est noté: $1R[1;\!2]$.

\paragraph{Se déplacer (\textit{D})}\phantom{}\\
Le personnage se déplace vers une tuile adjacente, puis les effets de celle-ci s’activent. Si la tuile de destination était face verso, elle est retournée dans le processus.

Cette action est désignée par la lettre $D$; le fait que le personnage $1$ se déplace vers la tuile de coordonnées $[1;\!2]$ est noté: $1D[1;\!2]$.

\paragraph{Pousser (\textit{P})}\textit{}\\
Cette action a le même effet que la précédente, au détail près qu’elle ne s’applique pas au personnage qui effectue l’action mais à un autre occupant de la salle où il se trouve. Les effets de la destination s’appliquent toujours, ce qui fait qu’un personnage peut en tuer un autre se trouvant sur la même tuile en le poussant dans une case mortelle adjacente. Une règle particulière (qui sera très importante par la suite) stipule qu’il est interdit de pousser depuis la tuile $\D$.

Cette action est désignée par la lettre $P$; le fait que le personnage $1$ pousse le joueur $2$ vers la tuile de coordonnées $[1;\!2]$ est noté: $1P2[1;\!2]$.

\begin{figure}[htb]
	\centering
	\begin{tabular}{ccc}
		\begin{tikzpicture}[scale=.43]
			\begin{pgfonlayer}{nodelayer}
				\node [style=none] (0) at (-5, 5) {};
				\node [style=none] (1) at (5, -5) {};
				\node [style=none] (2) at (-1, 1) {};
				\node [style=none] (3) at (1, -1) {};
				\node [style=none] (12) at (0, 0) {$\D$};
				\node [style=none] (15) at (-3, 5) {};
				\node [style=none] (16) at (-1, 5) {};
				\node [style=none] (17) at (1, 5) {};
				\node [style=none] (18) at (3, 5) {};
				\node [style=none] (19) at (5, 5) {};
				\node [style=none] (20) at (-5, 3) {};
				\node [style=none] (21) at (-3, 3) {};
				\node [style=none] (22) at (-1, 3) {};
				\node [style=none] (23) at (1, 3) {};
				\node [style=none] (24) at (3, 3) {};
				\node [style=none] (25) at (5, 3) {};
				\node [style=none] (27) at (-5, 1) {};
				\node [style=none] (28) at (-3, 1) {};
				\node [style=none] (29) at (1, 1) {};
				\node [style=none] (30) at (3, 1) {};
				\node [style=none] (31) at (5, 1) {};
				\node [style=none] (32) at (5, -1) {};
				\node [style=none] (33) at (3, -1) {};
				\node [style=none] (34) at (-1, -1) {};
				\node [style=none] (35) at (-3, -1) {};
				\node [style=none] (36) at (-5, -1) {};
				\node [style=none] (37) at (-5, -3) {};
				\node [style=none] (38) at (-3, -3) {};
				\node [style=none] (39) at (-1, -3) {};
				\node [style=none] (40) at (1, -3) {};
				\node [style=none] (41) at (3, -3) {};
				\node [style=none] (42) at (5, -3) {};
				\node [style=none] (43) at (3, -5) {};
				\node [style=none] (44) at (1, -5) {};
				\node [style=none] (45) at (-1, -5) {};
				\node [style=none] (46) at (-3, -5) {};
				\node [style=none] (47) at (-5, -5) {};
				\node [style=none] (48) at (-4, 2) {I};
				\node [style=none] (49) at (-2, 2) {II};
				\node [style=none] (50) at (0, 2) {III};
				\node [style=none] (51) at (2, 2) {$\S$};
				\node [style=none] (52) at (4, 2) {V};
				\node [style=none] (53) at (1.5, 2.5) {1};
				\node [style=none] (54) at (2.5, 1.5) {2};
			\end{pgfonlayer}
			\begin{pgfonlayer}{edgelayer}
				\fill [color=gray] (0.center) rectangle (1.center);
				\fill [color=cyan] (2.center) rectangle (3.center);
				\fill [color=cyan] (23.center) rectangle (30.center);
				\draw (0.center) to (19.center);
				\draw (20.center) to (25.center);
				\draw (31.center) to (27.center);
				\draw (36.center) to (32.center);
				\draw (42.center) to (37.center);
				\draw (47.center) to (1.center);
				\draw (1.center) to (19.center);
				\draw (18.center) to (43.center);
				\draw (44.center) to (17.center);
				\draw (16.center) to (45.center);
				\draw (46.center) to (15.center);
				\draw (0.center) to (47.center);
			\end{pgfonlayer}
		\end{tikzpicture}
		&
		\begin{tikzpicture}[scale=.43]
			\begin{pgfonlayer}{nodelayer}
				\node [style=none] (0) at (-5, 5) {};
				\node [style=none] (1) at (5, -5) {};
				\node [style=none] (2) at (-1, 1) {};
				\node [style=none] (3) at (1, -1) {};
				\node [style=none] (12) at (0, 0) {$\D$};
				\node [style=none] (15) at (-3, 5) {};
				\node [style=none] (16) at (-1, 5) {};
				\node [style=none] (17) at (1, 5) {};
				\node [style=none] (18) at (3, 5) {};
				\node [style=none] (19) at (5, 5) {};
				\node [style=none] (20) at (-5, 3) {};
				\node [style=none] (21) at (-3, 3) {};
				\node [style=none] (22) at (-1, 3) {};
				\node [style=none] (23) at (1, 3) {};
				\node [style=none] (24) at (3, 3) {};
				\node [style=none] (25) at (5, 3) {};
				\node [style=none] (27) at (-5, 1) {};
				\node [style=none] (28) at (-3, 1) {};
				\node [style=none] (29) at (1, 1) {};
				\node [style=none] (30) at (3, 1) {};
				\node [style=none] (31) at (5, 1) {};
				\node [style=none] (32) at (5, -1) {};
				\node [style=none] (33) at (3, -1) {};
				\node [style=none] (34) at (-1, -1) {};
				\node [style=none] (35) at (-3, -1) {};
				\node [style=none] (36) at (-5, -1) {};
				\node [style=none] (37) at (-5, -3) {};
				\node [style=none] (38) at (-3, -3) {};
				\node [style=none] (39) at (-1, -3) {};
				\node [style=none] (40) at (1, -3) {};
				\node [style=none] (41) at (3, -3) {};
				\node [style=none] (42) at (5, -3) {};
				\node [style=none] (43) at (3, -5) {};
				\node [style=none] (44) at (1, -5) {};
				\node [style=none] (45) at (-1, -5) {};
				\node [style=none] (46) at (-3, -5) {};
				\node [style=none] (47) at (-5, -5) {};
				\node [style=none] (48) at (-2, 2) {I};
				\node [style=none] (49) at (0, 2) {II};
				\node [style=none] (50) at (2, 2) {III};
				\node [style=none] (51) at (4, 2) {$\S$};
				\node [style=none] (52) at (-4, 2) {V};
				\node [style=none] (53) at (3.5, 2.5) {1};
				\node [style=none] (54) at (4.5, 1.5) {2};
			\end{pgfonlayer}
			\begin{pgfonlayer}{edgelayer}
				\fill [color=gray] (0.center) rectangle (1.center);
				\fill [color=cyan] (2.center) rectangle (3.center);
				\fill [color=cyan] (24.center) rectangle (31.center);
				\draw (0.center) to (19.center);
				\draw (20.center) to (25.center);
				\draw (31.center) to (27.center);
				\draw (36.center) to (32.center);
				\draw (42.center) to (37.center);
				\draw (47.center) to (1.center);
				\draw (1.center) to (19.center);
				\draw (18.center) to (43.center);
				\draw (44.center) to (17.center);
				\draw (16.center) to (45.center);
				\draw (46.center) to (15.center);
				\draw (0.center) to (47.center);
			\end{pgfonlayer}
		\end{tikzpicture}
		&
		\begin{tikzpicture}[scale=.43]
			\begin{pgfonlayer}{nodelayer}
				\node [style=none] (0) at (-5, 5) {};
				\node [style=none] (1) at (5, -5) {};
				\node [style=none] (2) at (-1, 1) {};
				\node [style=none] (3) at (1, -1) {};
				\node [style=none] (12) at (0, 0) {$\D$};
				\node [style=none] (15) at (-3, 5) {};
				\node [style=none] (16) at (-1, 5) {};
				\node [style=none] (17) at (1, 5) {};
				\node [style=none] (18) at (3, 5) {};
				\node [style=none] (19) at (5, 5) {};
				\node [style=none] (20) at (-5, 3) {};
				\node [style=none] (21) at (-3, 3) {};
				\node [style=none] (22) at (-1, 3) {};
				\node [style=none] (23) at (1, 3) {};
				\node [style=none] (24) at (3, 3) {};
				\node [style=none] (25) at (5, 3) {};
				\node [style=none] (27) at (-5, 1) {};
				\node [style=none] (28) at (-3, 1) {};
				\node [style=none] (29) at (1, 1) {};
				\node [style=none] (30) at (3, 1) {};
				\node [style=none] (31) at (5, 1) {};
				\node [style=none] (32) at (5, -1) {};
				\node [style=none] (33) at (3, -1) {};
				\node [style=none] (34) at (-1, -1) {};
				\node [style=none] (35) at (-3, -1) {};
				\node [style=none] (36) at (-5, -1) {};
				\node [style=none] (37) at (-5, -3) {};
				\node [style=none] (38) at (-3, -3) {};
				\node [style=none] (39) at (-1, -3) {};
				\node [style=none] (40) at (1, -3) {};
				\node [style=none] (41) at (3, -3) {};
				\node [style=none] (42) at (5, -3) {};
				\node [style=none] (43) at (3, -5) {};
				\node [style=none] (44) at (1, -5) {};
				\node [style=none] (45) at (-1, -5) {};
				\node [style=none] (46) at (-3, -5) {};
				\node [style=none] (47) at (-5, -5) {};
				\node [style=none] (48) at (0, 2) {I};
				\node [style=none] (49) at (2, 2) {II};
				\node [style=none] (50) at (4, 2) {III};
				\node [style=none] (51) at (6, 2) {$\S$};
				\node [style=none] (52) at (-2, 2) {V};
				\node [style=none] (53) at (5.5, 2.5) {1};
				\node [style=none] (54) at (6.5, 1.5) {2};
				\node [style=none] (55) at (7, 1) {};
			\end{pgfonlayer}
			\begin{pgfonlayer}{edgelayer}
				\fill [color=gray] (0.center) rectangle (1.center);
				\fill [color=cyan] (2.center) rectangle (3.center);
				\fill [color=cyan] (25.center) rectangle (55.center);
				\fill [color=white] (20.center) rectangle (28.center);
				\draw (25.center) rectangle (55.center);
				\draw (0.center) to (19.center);
				\draw (20.center) to (25.center);
				\draw (31.center) to (27.center);
				\draw (36.center) to (32.center);
				\draw (42.center) to (37.center);
				\draw (1.center) to (19.center);
				\draw (18.center) to (43.center);
				\draw (44.center) to (17.center);
				\draw (16.center) to (45.center);
				\draw (46.center) to (15.center);
				\draw (0.center) to (47.center);
				\draw [color=white] (20.center) to (27.center);
			\end{pgfonlayer}
		\end{tikzpicture}
		\\
		avant & après un $C$ & après un second $C$
	\end{tabular}
	\caption{Illustration de deux actions $C$. La première effectue un décalage cyclique, et la seconde est une fin de partie (victorieuse) parce que $\S$ sort du cadre en emportant tous les personnages.}
	\label{fig_ctrl}
\end{figure}
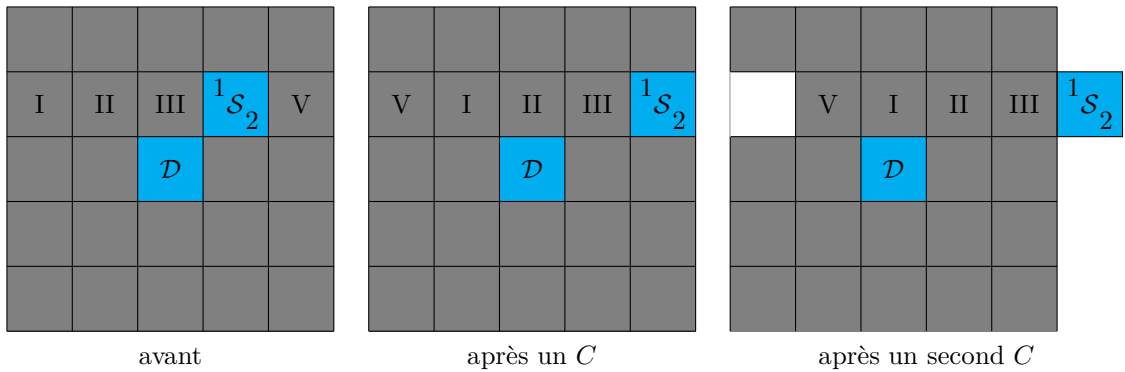

\paragraph{Contrôler (\textit{C})}\textit{}\\
Le personnage décale cycliquement d’un cran toutes les tuiles de la ligne ou la colonne où il se situe, dans le sens qu’il souhaite (voir \cref{fig_ctrl}). Tous les personnages se déplacent avec la tuile qu’ils occupent. La tuile $\D$ devant toujours rester en $[0;\!0]$, il est impossible de décaler les rangées $[0;]$ et $[;\!0]$, et par suite l’action n’a aucun effet si le personnage l’effectuant se trouve en $\D$. Autre subtilité: il est possible de décaler plusieurs fois une rangée dans un tour, mais seulement dans le même sens. Enfin, si tous les joueurs se trouvent en $\S$ et qu’une action \textit{contrôler} la décale hors du plateau, la partie est remportée.\footnote{Thématiquement, les candidats sortent du complexe, remportant ainsi le jeu.}

La salle de contrôle $\C$ a intuitivement pour effet de donner une action $C$ valable n’importe où sur le plateau. Plus précisément, chaque fois qu’un personnage entre dans $\C$ son contrôleur choisit n’importe quelle rangée et la déplace comme s’il effectuait l’action $C$ dessus.

Cette action est désignée par la lettre $C$; le fait que le personnage $1$ décale la colonne $[2;]$ vers le haut est noté: $1P\!\uparrow\![2;]$. Lorsque le décalage est dû à $\C$, l’on écrit seulement la direction et la coordonnée à la suite de l’action. Par exemple si le personnage $1$ se déplace vers la tuile de coordonnées $[1;\!2]$ qui s’avère être $\C$ et qu’il décale la colonne $[2;]$ vers le haut, l’on note: $1D[1;\!2]\!\uparrow\![2;]$.\\

\begin{figure}[htb]
	\centering
	\begin{tabular}{cc}
		\begin{tikzpicture}[scale=.5]
			\begin{pgfonlayer}{nodelayer}
				\node [style=none] (0) at (-5, 5) {};
				\node [style=none] (1) at (5, -5) {};
				\node [style=none] (2) at (-1, 1) {};
				\node [style=none] (3) at (1, -1) {};
				\node [style=none] (12) at (0, 0) {$\D$};
				\node [style=none] (15) at (-3, 5) {};
				\node [style=none] (16) at (-1, 5) {};
				\node [style=none] (17) at (1, 5) {};
				\node [style=none] (18) at (3, 5) {};
				\node [style=none] (19) at (5, 5) {};
				\node [style=none] (20) at (-5, 3) {};
				\node [style=none] (21) at (-3, 3) {};
				\node [style=none] (22) at (-1, 3) {};
				\node [style=none] (23) at (1, 3) {};
				\node [style=none] (24) at (3, 3) {};
				\node [style=none] (25) at (5, 3) {};
				\node [style=none] (27) at (-5, 1) {};
				\node [style=none] (28) at (-3, 1) {};
				\node [style=none] (29) at (1, 1) {};
				\node [style=none] (30) at (3, 1) {};
				\node [style=none] (31) at (5, 1) {};
				\node [style=none] (32) at (5, -1) {};
				\node [style=none] (33) at (3, -1) {};
				\node [style=none] (34) at (-1, -1) {};
				\node [style=none] (35) at (-3, -1) {};
				\node [style=none] (36) at (-5, -1) {};
				\node [style=none] (37) at (-5, -3) {};
				\node [style=none] (38) at (-3, -3) {};
				\node [style=none] (39) at (-1, -3) {};
				\node [style=none] (40) at (1, -3) {};
				\node [style=none] (41) at (3, -3) {};
				\node [style=none] (42) at (5, -3) {};
				\node [style=none] (43) at (3, -5) {};
				\node [style=none] (44) at (1, -5) {};
				\node [style=none] (45) at (-1, -5) {};
				\node [style=none] (46) at (-3, -5) {};
				\node [style=none] (47) at (-5, -5) {};
				\node [style=none] (48) at (2, 0) {};
				\node [style=none] (49) at (2, 0) {$\V$};
				\node [style=none] (50) at (1.5, 0.5) {1};
				\node [style=none] (51) at (2, 0.5) {2};
				\node [style=none] (52) at (2.5, 0.5) {3};
			\end{pgfonlayer}
			\begin{pgfonlayer}{edgelayer}
				\fill [color=gray] (0.center) rectangle (1.center);
				\fill [color=cyan] (2.center) rectangle (3.center);
				\fill [color=green] (29.center) rectangle (33.center);
				\draw (0.center) to (19.center);
				\draw (20.center) to (25.center);
				\draw (31.center) to (27.center);
				\draw (36.center) to (32.center);
				\draw (42.center) to (37.center);
				\draw (47.center) to (1.center);
				\draw (1.center) to (19.center);
				\draw (18.center) to (43.center);
				\draw (44.center) to (17.center);
				\draw (16.center) to (45.center);
				\draw (46.center) to (15.center);
				\draw (0.center) to (47.center);
			\end{pgfonlayer}
		\end{tikzpicture}
		&
		\begin{tikzpicture}[scale=.5]
			\begin{pgfonlayer}{nodelayer}
				\node [style=none] (0) at (-5, 5) {};
				\node [style=none] (1) at (5, -5) {};
				\node [style=none] (2) at (-1, 1) {};
				\node [style=none] (3) at (1, -1) {};
				\node [style=none] (12) at (0, 0) {$\D$};
				\node [style=none] (15) at (-3, 5) {};
				\node [style=none] (16) at (-1, 5) {};
				\node [style=none] (17) at (1, 5) {};
				\node [style=none] (18) at (3, 5) {};
				\node [style=none] (19) at (5, 5) {};
				\node [style=none] (20) at (-5, 3) {};
				\node [style=none] (21) at (-3, 3) {};
				\node [style=none] (22) at (-1, 3) {};
				\node [style=none] (23) at (1, 3) {};
				\node [style=none] (24) at (3, 3) {};
				\node [style=none] (25) at (5, 3) {};
				\node [style=none] (27) at (-5, 1) {};
				\node [style=none] (28) at (-3, 1) {};
				\node [style=none] (29) at (1, 1) {};
				\node [style=none] (30) at (3, 1) {};
				\node [style=none] (31) at (5, 1) {};
				\node [style=none] (32) at (5, -1) {};
				\node [style=none] (33) at (3, -1) {};
				\node [style=none] (34) at (-1, -1) {};
				\node [style=none] (35) at (-3, -1) {};
				\node [style=none] (36) at (-5, -1) {};
				\node [style=none] (37) at (-5, -3) {};
				\node [style=none] (38) at (-3, -3) {};
				\node [style=none] (39) at (-1, -3) {};
				\node [style=none] (40) at (1, -3) {};
				\node [style=none] (41) at (3, -3) {};
				\node [style=none] (42) at (5, -3) {};
				\node [style=none] (43) at (3, -5) {};
				\node [style=none] (44) at (1, -5) {};
				\node [style=none] (45) at (-1, -5) {};
				\node [style=none] (46) at (-3, -5) {};
				\node [style=none] (47) at (-5, -5) {};
				\node [style=none] (48) at (2, 0) {};
				\node [style=none] (49) at (2, 0) {$\V$};
				\node [style=none] (51) at (1.5, 0.5) {2};
				\node [style=none] (52) at (3.5, 0.5) {1};
				\node [style=none] (53) at (-0.5, 0.5) {3};
				\node [style=none] (54) at (4, 0) {$\V$};
			\end{pgfonlayer}
			\begin{pgfonlayer}{edgelayer}
				\fill [color=gray] (0.center) rectangle (1.center);
				\fill [color=cyan] (2.center) rectangle (3.center);
				\fill [color=green] (29.center) rectangle (33.center);
				\fill [color=green] (30.center) rectangle (32.center);
				\draw (0.center) to (19.center);
				\draw (20.center) to (25.center);
				\draw (31.center) to (27.center);
				\draw (36.center) to (32.center);
				\draw (42.center) to (37.center);
				\draw (47.center) to (1.center);
				\draw (1.center) to (19.center);
				\draw (18.center) to (43.center);
				\draw (44.center) to (17.center);
				\draw (16.center) to (45.center);
				\draw (46.center) to (15.center);
				\draw (0.center) to (47.center);
			\end{pgfonlayer}
		\end{tikzpicture}
		\\
		avant & après
	\end{tabular}
	\caption{Illustration des actions $D$ et $P$. Le personnage 1 se déplace vers $[2;\!0]$ et le 2 pousse le 3 vers $[0;\!0]$.}
	\label{fig_depouss}
\end{figure}
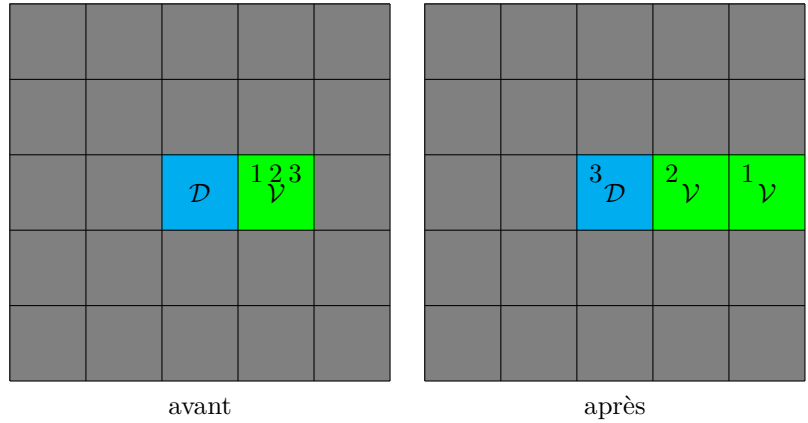

La programmation se note en donnant pour chaque joueur la ou les actions choisies, séparés par des |. Ensuite, son exécution se décrit comme expliqué ci-dessus, là encore séparé par des |. Les personnages effectuent leur première action  dans un ordre qui évolue au fil de la partie\footnote{Au premier tour 123… puis 234…1 au second etc.}, puis leur seconde dans ce même ordre. Si un personnage n’a qu’une seule action de programmée, son contrôleur choisit s’il l’effectue lors du premier ou de second passage. Toute action programmée doit être exécutée si possible, même si cela entraîne la mort du personnage; le joueur a seulement le choix des paramètres de l’action.

\begin{exmp}\label{ex1}
	Supposons qu’à la fin du premier tour, tous les personnages se trouvent en $[1;\!0]$, qui est un $\V$, et que leurs actions $R$ leur aient révélé que $[-1;\!0]$ et $[2;\!0]$ sont également des $\V$. Ils effectuent la programmation suivante:\vspace{-3mm}
	\[\mathbf{2:}2DR\sep3PR\sep1DR\]\vspace{-7mm}
	
	Ils choisissent alors de commencer l’exécution du tour par:\vspace{-3mm}
	\[\mathbf{2:}2D[2;\!0]\sep3P1[0;\!0]\sep1D[-1;\!0]\sep\dots\]\vspace{-7mm}
	
	ce qui leur permet d’être dispersés sur plusieurs tuiles différentes pour avoir davantage de choix dans les tuiles à regarder.
\end{exmp}

\begin{figure}[htb]
	\centering
	\begin{tabular}{cc}
		\begin{tikzpicture}[scale=.5]
			\begin{pgfonlayer}{nodelayer}
				\node [style=none] (0) at (-5, 5) {};
				\node [style=none] (1) at (5, -5) {};
				\node [style=none] (2) at (-1, 1) {};
				\node [style=none] (3) at (1, -1) {};
				\node [style=none] (12) at (0, 0) {$\D$};
				\node [style=none] (15) at (-3, 5) {};
				\node [style=none] (16) at (-1, 5) {};
				\node [style=none] (17) at (1, 5) {};
				\node [style=none] (18) at (3, 5) {};
				\node [style=none] (19) at (5, 5) {};
				\node [style=none] (20) at (-5, 3) {};
				\node [style=none] (21) at (-3, 3) {};
				\node [style=none] (22) at (-1, 3) {};
				\node [style=none] (23) at (1, 3) {};
				\node [style=none] (24) at (3, 3) {};
				\node [style=none] (25) at (5, 3) {};
				\node [style=none] (27) at (-5, 1) {};
				\node [style=none] (28) at (-3, 1) {};
				\node [style=none] (29) at (1, 1) {};
				\node [style=none] (30) at (3, 1) {};
				\node [style=none] (31) at (5, 1) {};
				\node [style=none] (32) at (5, -1) {};
				\node [style=none] (33) at (3, -1) {};
				\node [style=none] (34) at (-1, -1) {};
				\node [style=none] (35) at (-3, -1) {};
				\node [style=none] (36) at (-5, -1) {};
				\node [style=none] (37) at (-5, -3) {};
				\node [style=none] (38) at (-3, -3) {};
				\node [style=none] (39) at (-1, -3) {};
				\node [style=none] (40) at (1, -3) {};
				\node [style=none] (41) at (3, -3) {};
				\node [style=none] (42) at (5, -3) {};
				\node [style=none] (43) at (3, -5) {};
				\node [style=none] (44) at (1, -5) {};
				\node [style=none] (45) at (-1, -5) {};
				\node [style=none] (46) at (-3, -5) {};
				\node [style=none] (47) at (-5, -5) {};
				\node [style=none] (48) at (2, 0) {};
				\node [style=none] (49) at (2, 0) {$\V$};
				\node [style=none] (50) at (1.5, 0.5) {1};
				\node [style=none] (51) at (2, 0.5) {2};
				\node [style=none] (52) at (2.5, 0.5) {3};
			\end{pgfonlayer}
			\begin{pgfonlayer}{edgelayer}
				\fill [color=gray] (0.center) rectangle (1.center);
				\fill [color=cyan] (2.center) rectangle (3.center);
				\fill [color=green] (29.center) rectangle (33.center);
				\draw (0.center) to (19.center);
				\draw (20.center) to (25.center);
				\draw (31.center) to (27.center);
				\draw (36.center) to (32.center);
				\draw (42.center) to (37.center);
				\draw (47.center) to (1.center);
				\draw (1.center) to (19.center);
				\draw (18.center) to (43.center);
				\draw (44.center) to (17.center);
				\draw (16.center) to (45.center);
				\draw (46.center) to (15.center);
				\draw (0.center) to (47.center);
			\end{pgfonlayer}
		\end{tikzpicture}
		&
		\begin{tikzpicture}[scale=.5]
			\begin{pgfonlayer}{nodelayer}
				\node [style=none] (0) at (-5, 5) {};
				\node [style=none] (1) at (5, -5) {};
				\node [style=none] (2) at (-1, 1) {};
				\node [style=none] (3) at (1, -1) {};
				\node [style=none] (12) at (0, 0) {$\D$};
				\node [style=none] (15) at (-3, 5) {};
				\node [style=none] (16) at (-1, 5) {};
				\node [style=none] (17) at (1, 5) {};
				\node [style=none] (18) at (3, 5) {};
				\node [style=none] (19) at (5, 5) {};
				\node [style=none] (20) at (-5, 3) {};
				\node [style=none] (21) at (-3, 3) {};
				\node [style=none] (22) at (-1, 3) {};
				\node [style=none] (23) at (1, 3) {};
				\node [style=none] (24) at (3, 3) {};
				\node [style=none] (25) at (5, 3) {};
				\node [style=none] (27) at (-5, 1) {};
				\node [style=none] (28) at (-3, 1) {};
				\node [style=none] (29) at (1, 1) {};
				\node [style=none] (30) at (3, 1) {};
				\node [style=none] (31) at (5, 1) {};
				\node [style=none] (32) at (5, -1) {};
				\node [style=none] (33) at (3, -1) {};
				\node [style=none] (34) at (-1, -1) {};
				\node [style=none] (35) at (-3, -1) {};
				\node [style=none] (36) at (-5, -1) {};
				\node [style=none] (37) at (-5, -3) {};
				\node [style=none] (38) at (-3, -3) {};
				\node [style=none] (39) at (-1, -3) {};
				\node [style=none] (40) at (1, -3) {};
				\node [style=none] (41) at (3, -3) {};
				\node [style=none] (42) at (5, -3) {};
				\node [style=none] (43) at (3, -5) {};
				\node [style=none] (44) at (1, -5) {};
				\node [style=none] (45) at (-1, -5) {};
				\node [style=none] (46) at (-3, -5) {};
				\node [style=none] (47) at (-5, -5) {};
				\node [style=none] (48) at (2, 0) {};
				\node [style=none] (49) at (2, 0) {$\V$};
				\node [style=none] (51) at (1.5, 0.5) {3};
				\node [style=none] (52) at (3.5, 0.5) {2};
				\node [style=none] (53) at (-2.5, 0.5) {1};
				\node [style=none] (54) at (4, 0) {$\V$};
				\node [style=none] (55) at (-2, 0) {$\V$};
			\end{pgfonlayer}
			\begin{pgfonlayer}{edgelayer}
				\fill [color=gray] (0.center) rectangle (1.center);
				\fill [color=cyan] (2.center) rectangle (3.center);
				\fill [color=green] (29.center) rectangle (33.center);
				\fill [color=green] (30.center) rectangle (32.center);
				\fill [color=green] (28.center) rectangle (34.center);
				\draw (0.center) to (19.center);
				\draw (20.center) to (25.center);
				\draw (31.center) to (27.center);
				\draw (36.center) to (32.center);
				\draw (42.center) to (37.center);
				\draw (47.center) to (1.center);
				\draw (1.center) to (19.center);
				\draw (18.center) to (43.center);
				\draw (44.center) to (17.center);
				\draw (16.center) to (45.center);
				\draw (46.center) to (15.center);
				\draw (0.center) to (47.center);
			\end{pgfonlayer}
		\end{tikzpicture}
		\\
		avant & après	
	\end{tabular}
	\caption{Illustration de l’\cref{ex1}. Rappelons l’exécution du tour:\\
		\phantom{pour meubler} $\mathbf{2:}2D[2;\!0]\sep3P1[0;\!0]\sep1D[-1;\!0]\sep\dots$}
	\label{fig_ex1}
\end{figure}
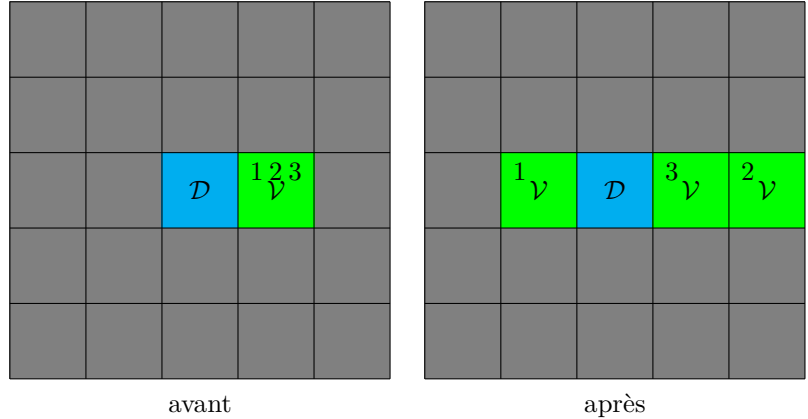

\section{Parties optimales}\label{sec_opti}

Nous pouvons à présent étudier les parties optimales à proprement parler. Nous démontrerons d’abord qu’il est impossible de gagner en un tour, puis qu’il est au contraire possible de le faire en deux, et enfin qu’à condition de changer légèrement les règles, la victoire au premier tour devient accessible. Notons que ce que nous appelons \og victoire \fg{} est celle du mode solo ou la \og victoire complète \fg{} du mode coopération, c’est-à-dire qu’aucune mort n’est tolérée et de fait que tous les personnages doivent impérativement se trouver en $\S$ à la fin de la partie.

\subsection{Borne inférieure}

Nous cherchons à établir la proposition suivante.

\begin{prop}\label{prop_binf}
	Quelque soit le nombre de personnages, il est impossible de remporter une partie solo ou en coopération en un seul tour.
\end{prop}

Le résultat s’appuie d’une part sur la règle selon laquelle le seul moyen de quitter $\D$ est d’utiliser l’action de déplacement, car pousser y est interdit, et d’autre part sur le lemme suivant. L’on fixe une partie victorieuse (qui fait potentiellement plusieurs tours), et l’on note $\Pi$ l’ensemble des personnages qui ont atteint une salle non-bleue à un moment de la partie.

\begin{lem}\label{lem_binf1}
	Le dernier membre de $\Pi$ à entrer (pour la dernière fois) dans la salle $\S$ le fait forcément par l’action de déplacement.
\end{lem}

\begin{dem}
	Soit $p\in\Pi$ un personnage qui entre pour la dernière fois en $\S$ autrement que par l’action $D$. Le seul autre moyen que $p$ ait de changer de salle étant que quelqu’un le pousse, il y a donc un autre personnage $p’$ qui l’a poussé depuis une salle $\X$ qui n’était pas $\S$ (pour être poussé depuis $\S$, $p$ aurait dû déjà s’y trouver) ni $\C$ (il est interdit de pousser depuis $\C$). Ainsi, puisque $p’$ était en $\X$, $p’\in\Pi$, et puisqu’il y était toujours après avoir poussé $p$, ce dernier n’est pas le dernier membre de $\Pi$ à rejoindre $\S$.
\end{dem}

En invoquant de nouveau la règle interdisant de pousser depuis $\D$, l’on peut raffiner ce premier lemme en un second.

\begin{lem}\label{lem_binf2}
	Si $\Pi\neq\emptyset$, alors il existe un $p\in\Pi$ ayant utilisé au moins deux fois l’action de déplacement.
\end{lem}

\begin{dem}
	Soit $p$ le dernier membre de $\Pi$ à rejoindre $\S$. D’après le \cref{lem_binf1}, il a utilisé une action $D$ pour finalement rejoindre $\S$. En outre, $p\in\Pi$ signifie qu’il a atteint une case non-bleue. Or il est parti comme tout les autres personnages de $\D$ et n’a pu en sortir pour atteindre une case non-bleue que par une action $D$, distincte de celle garantie par le \cref{lem_binf1}. En effet, après ce dernier déplacement, $p$ reste en $\S$ jusqu’à la fin de la partie; s’il était confondu avec le premier déplacement, $p$ n’aurait vu aucune autre tuile que $\D$ et $\S$ et ne serait donc pas dans $\Pi$.
\end{dem}

\vspace{-11pt}Nous pouvons à présent terminer la preuve.

\begin{dem}[ (de la \cref{prop_binf})]
	Supposons par l’absurde une partie remportée en un tour. Puisqu’aucun personnage ne peut effectuer deux actions $D$ dans le même tour, le \cref{lem_binf2} implique que $\Pi=\emptyset$, c’est-à-dire qu’aucun personnage n’a atteint une tuile non-bleue. Or ceci est impossible $\S$ ne peut pas être adjacente à $\D$ au début de la partie, et qu’aucune tuile ne peut être déplacée si tous les joueurs restent en $\D$.
\end{dem}

\begin{nb}
	Une victoire partielle n’est en fait pas possible non plus, parce qu’il faudrait pour cela que tous les membres de $\Pi$ atteignent $\S$ sauf un qui doit mourir avant que les autres n’essaient de décaler $\S$ hors du plateau. L’on peut alors conclure par une analyse de cas des manières de tuer un personnage donnée en \cref{anx}.
\end{nb}

\subsection{Ouverture véloce}

Puisque la partie ne peut pas être remportée en un seul tour, nous cherchant à présent une victoire en deux tours.

\begin{defn}
	L’ouverture véloce est définie comme suit:
	\begin{itemize}
		\item à 1 personnage: $\mathbf{1:}1DC$;
		\item à $n>1$ personnages : $\mathbf{1:}1RD\sep\dots\sep(n-2)RD\sep(n-1)DC\sep nDR$.
	\end{itemize}
\end{defn}

\begin{defn}
	Les joueurs sont dits \textit{V-chanceux} s’ils choisissent un repère $\mathfrak{R}’$ obtenu par rotation et/ou réflexion (\textit{id est} les symétries du plateau) du repère standard qui soit tel que:
	\begin{itemize}
		\item $\S$ soit en $[2;\!1]$ selon $\mathfrak{R}’$;
		\item la tuile en $[1;\!0]$ selon $\mathfrak{R}’$ soit une salle vide, une salle noire ou un tunnel.
	\end{itemize}
\end{defn}

Ces trois tuiles sont les seules de la distribution solo et coop standard à ne pas poser problème pour la stratégie que l’on essaie d’appliquer. Le tunnel n’a aucun effet tant que l’on n’en visite pas un autre et la salle noire empêche seulement l’action $R$. Le repère $\mathfrak{R}’$ est uniquement un moyen de formaliser le fait que par symétrie du plateau, il est possible de partir dans n’importe quelle direction.

\begin{prop}
	L’ouverture véloce permet de gagner en deux tours si les joueurs sont V-chanceux.
\end{prop}

\begin{dem}
	Quitte à appliquer au plateau les rotations et réflexions nécessaires, l’on peut supposer que $\mathfrak{R}’$ n’est autre que le repère standard. La condition de V-chance est alors résumée par la \cref{fig_v1}(a), avec $\X$ une case qui n’est pas nécessairement verte mais qui ne trouble pas la stratégie comme expliqué plus haut. L’on peut se restreindre au cas où il n’y a qu’un seul personnage, puisque les autres ne font que regarder aux alentours et imiter les déplacements du seul personnage ayant programmé $DC$.
	
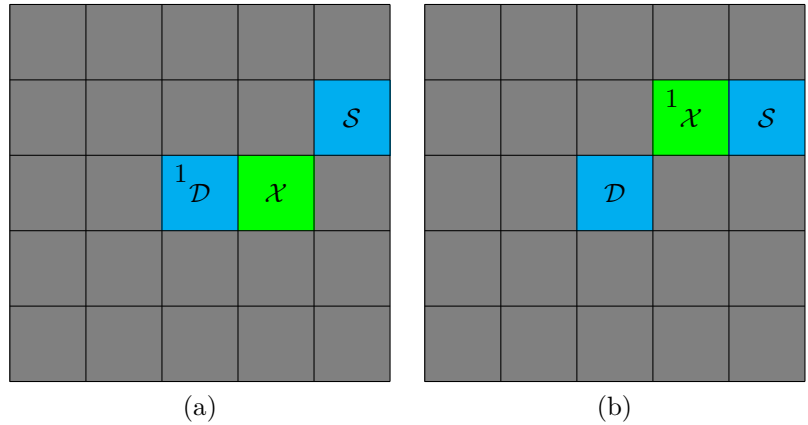
\begin{figure}[htb]
	\centering
	\begin{tabular}{cc}
		\begin{tikzpicture}[scale=.5]
			\begin{pgfonlayer}{nodelayer}
				\node [style=none] (0) at (-5, 5) {};
				\node [style=none] (1) at (5, -5) {};
				\node [style=none] (2) at (-1, 1) {};
				\node [style=none] (3) at (1, -1) {};
				\node [style=none] (12) at (0, 0) {$\D$};
				\node [style=none] (15) at (-3, 5) {};
				\node [style=none] (16) at (-1, 5) {};
				\node [style=none] (17) at (1, 5) {};
				\node [style=none] (18) at (3, 5) {};
				\node [style=none] (19) at (5, 5) {};
				\node [style=none] (20) at (-5, 3) {};
				\node [style=none] (21) at (-3, 3) {};
				\node [style=none] (22) at (-1, 3) {};
				\node [style=none] (23) at (1, 3) {};
				\node [style=none] (24) at (3, 3) {};
				\node [style=none] (25) at (5, 3) {};
				\node [style=none] (27) at (-5, 1) {};
				\node [style=none] (28) at (-3, 1) {};
				\node [style=none] (29) at (1, 1) {};
				\node [style=none] (30) at (3, 1) {};
				\node [style=none] (31) at (5, 1) {};
				\node [style=none] (32) at (5, -1) {};
				\node [style=none] (33) at (3, -1) {};
				\node [style=none] (34) at (-1, -1) {};
				\node [style=none] (35) at (-3, -1) {};
				\node [style=none] (36) at (-5, -1) {};
				\node [style=none] (37) at (-5, -3) {};
				\node [style=none] (38) at (-3, -3) {};
				\node [style=none] (39) at (-1, -3) {};
				\node [style=none] (40) at (1, -3) {};
				\node [style=none] (41) at (3, -3) {};
				\node [style=none] (42) at (5, -3) {};
				\node [style=none] (43) at (3, -5) {};
				\node [style=none] (44) at (1, -5) {};
				\node [style=none] (45) at (-1, -5) {};
				\node [style=none] (46) at (-3, -5) {};
				\node [style=none] (47) at (-5, -5) {};
				\node [style=none] (48) at (2, 0) {};
				\node [style=none] (49) at (2, 0) {$\X$};
				\node [style=none] (50) at (-0.5, 0.5) {1};
				\node [style=none] (51) at (4, 2) {$\S$};
			\end{pgfonlayer}
			\begin{pgfonlayer}{edgelayer}
				\fill [color=gray] (0.center) rectangle (1.center);
				\fill [color=cyan] (2.center) rectangle (3.center);
				\fill [color=green] (29.center) rectangle (33.center);
				\fill [color=cyan] (24.center) rectangle (31.center);
				\draw (0.center) to (19.center);
				\draw (20.center) to (25.center);
				\draw (31.center) to (27.center);
				\draw (36.center) to (32.center);
				\draw (42.center) to (37.center);
				\draw (47.center) to (1.center);
				\draw (1.center) to (19.center);
				\draw (18.center) to (43.center);
				\draw (44.center) to (17.center);
				\draw (16.center) to (45.center);
				\draw (46.center) to (15.center);
				\draw (0.center) to (47.center);
			\end{pgfonlayer}
		\end{tikzpicture}
		&
		\begin{tikzpicture}[scale=.5]
			\begin{pgfonlayer}{nodelayer}
				\node [style=none] (0) at (-5, 5) {};
				\node [style=none] (1) at (5, -5) {};
				\node [style=none] (2) at (-1, 1) {};
				\node [style=none] (3) at (1, -1) {};
				\node [style=none] (12) at (0, 0) {$\D$};
				\node [style=none] (15) at (-3, 5) {};
				\node [style=none] (16) at (-1, 5) {};
				\node [style=none] (17) at (1, 5) {};
				\node [style=none] (18) at (3, 5) {};
				\node [style=none] (19) at (5, 5) {};
				\node [style=none] (20) at (-5, 3) {};
				\node [style=none] (21) at (-3, 3) {};
				\node [style=none] (22) at (-1, 3) {};
				\node [style=none] (23) at (1, 3) {};
				\node [style=none] (24) at (3, 3) {};
				\node [style=none] (25) at (5, 3) {};
				\node [style=none] (27) at (-5, 1) {};
				\node [style=none] (28) at (-3, 1) {};
				\node [style=none] (29) at (1, 1) {};
				\node [style=none] (30) at (3, 1) {};
				\node [style=none] (31) at (5, 1) {};
				\node [style=none] (32) at (5, -1) {};
				\node [style=none] (33) at (3, -1) {};
				\node [style=none] (34) at (-1, -1) {};
				\node [style=none] (35) at (-3, -1) {};
				\node [style=none] (36) at (-5, -1) {};
				\node [style=none] (37) at (-5, -3) {};
				\node [style=none] (38) at (-3, -3) {};
				\node [style=none] (39) at (-1, -3) {};
				\node [style=none] (40) at (1, -3) {};
				\node [style=none] (41) at (3, -3) {};
				\node [style=none] (42) at (5, -3) {};
				\node [style=none] (43) at (3, -5) {};
				\node [style=none] (44) at (1, -5) {};
				\node [style=none] (45) at (-1, -5) {};
				\node [style=none] (46) at (-3, -5) {};
				\node [style=none] (47) at (-5, -5) {};
				\node [style=none] (49) at (2, 2) {$\X$};
				\node [style=none] (50) at (1.5, 2.5) {1};
				\node [style=none] (51) at (4, 2) {$\S$};
			\end{pgfonlayer}
			\begin{pgfonlayer}{edgelayer}
				\fill [color=gray] (0.center) rectangle (1.center);
				\fill [color=cyan] (2.center) rectangle (3.center);
				\fill [color=green] (23.center) rectangle (30.center);
				\fill [color=cyan] (24.center) rectangle (31.center);
				\draw (0.center) to (19.center);
				\draw (20.center) to (25.center);
				\draw (31.center) to (27.center);
				\draw (36.center) to (32.center);
				\draw (42.center) to (37.center);
				\draw (47.center) to (1.center);
				\draw (1.center) to (19.center);
				\draw (18.center) to (43.center);
				\draw (44.center) to (17.center);
				\draw (16.center) to (45.center);
				\draw (46.center) to (15.center);
				\draw (0.center) to (47.center);
			\end{pgfonlayer}
		\end{tikzpicture}
		\\
		(a) & (b)	
	\end{tabular}
	\caption{Illustration de la partie optimale: (a) configuration initiale, (b) fin du premier tour.}
	\label{fig_v1}
\end{figure}

	Lorsque le plateau est ainsi formé et que le personnage unique a programmé $\mathbf{1:}1DC$, il peut alors dérouler la partie suivante (le $\#$ représente une partie remportée):
	\begin{align*}
		\mathbf{1:} & 1D[1;\!0] \sep 1C\!\uparrow\![1;]\\
		\mathbf{2:} & 1D[2;\!1] \sep 1C\!\to\![;1]\#\\
	\end{align*}
	\phantom{text}\vspace{-15mm}\\
\end{dem}

\subsection{Ouverture téméraire}

La preuve de l’impossibilité de finir en un seul tour reposait fortement sur l’interdiction de pousser depuis $\D$. Il est naturel de se demander ce qu’il advient si l’on retire cette règle. Comme nous le verrons il devient alors possible de le faire avec six personnages si le complexe contient une salle de contrôle (notée $\C$). Cette salle ne fait pas partie de la composition recommandée pour les modes solo et coopération, mais le livret de règles lui-même nous invite à modifier la composition de départ du complexe. Nous considérerons donc la composition recommandée dans laquelle l’on remplace une des salles vides par la salle de contrôle.

\begin{defn}
	L’ouverture téméraire est une ouverture à six personnages définie comme suit:
	\[\mathbf{1:} 1PD\sep 2PD\sep 3CD\sep 4DC\sep 5DC\sep 6DC\]
\end{defn}

\begin{defn}
	Les joueurs sont dits \textit{T-chanceux} s’ils sont V-chanceux à la différence près que la tuile en $[1;\!0]$ selon $\mathfrak{R}’$ doit être la salle de contrôle $\C$.
\end{defn}

\begin{prop}
	L’ouverture téméraire permet de gagner en deux tours si les joueurs sont T-chanceux.
\end{prop}

\begin{dem}
	Là encore, l’on suppose par symétrie que $\mathfrak{R}’$ est le repère ordinaire. L’on peut alors dérouler la partie suivante à six joueurs, illustrée en \cref{fig_t1}.\vspace{-3mm}
	
	\begin{align*}
		\mathbf{1:} & 1P3[1;\!0]\!\leftarrow\![;1]\sep 2P4[0;\!1]\!\leftarrow\![;1]\overset{\color{red} (b)}{\sep}
		3C\!\uparrow\![1;] \overset{\color{red} (c)}{\sep} 4D[1;0] \sep 5D[1;0] \sep 6D[1;0] \sep\\
		& 1D[1;0] \sep 2D[1;0] \sep 3D[1;0] \overset{\color{red} (d)}{\sep} 4C\!\leftarrow\![;1]\sep 5\!\leftarrow\![;1]\sep 6\!\leftarrow\![;1]\sep
	\end{align*}
	
	\begin{figure}[htb]
	\centering
	\begin{tabular}{cc}
		\begin{tikzpicture}[scale=.5]
			\begin{pgfonlayer}{nodelayer}
				\node [style=none] (0) at (-5, 5) {};
				\node [style=none] (1) at (5, -5) {};
				\node [style=none] (2) at (-1, 1) {};
				\node [style=none] (3) at (1, -1) {};
				\node [style=none] (12) at (0, 0) {$\D$};
				\node [style=none] (15) at (-3, 5) {};
				\node [style=none] (16) at (-1, 5) {};
				\node [style=none] (17) at (1, 5) {};
				\node [style=none] (18) at (3, 5) {};
				\node [style=none] (19) at (5, 5) {};
				\node [style=none] (20) at (-5, 3) {};
				\node [style=none] (21) at (-3, 3) {};
				\node [style=none] (22) at (-1, 3) {};
				\node [style=none] (23) at (1, 3) {};
				\node [style=none] (24) at (3, 3) {};
				\node [style=none] (25) at (5, 3) {};
				\node [style=none] (27) at (-5, 1) {};
				\node [style=none] (28) at (-3, 1) {};
				\node [style=none] (29) at (1, 1) {};
				\node [style=none] (30) at (3, 1) {};
				\node [style=none] (31) at (5, 1) {};
				\node [style=none] (32) at (5, -1) {};
				\node [style=none] (33) at (3, -1) {};
				\node [style=none] (34) at (-1, -1) {};
				\node [style=none] (35) at (-3, -1) {};
				\node [style=none] (36) at (-5, -1) {};
				\node [style=none] (37) at (-5, -3) {};
				\node [style=none] (38) at (-3, -3) {};
				\node [style=none] (39) at (-1, -3) {};
				\node [style=none] (40) at (1, -3) {};
				\node [style=none] (41) at (3, -3) {};
				\node [style=none] (42) at (5, -3) {};
				\node [style=none] (43) at (3, -5) {};
				\node [style=none] (44) at (1, -5) {};
				\node [style=none] (45) at (-1, -5) {};
				\node [style=none] (46) at (-3, -5) {};
				\node [style=none] (47) at (-5, -5) {};
				\node [style=none] (48) at (2, 0) {};
				\node [style=none] (49) at (2, 0) {$\M$};
				\node [style=none] (50) at (-0.5, 0.5) {1};
				\node [style=none] (51) at (4, 2) {$\S$};
				\node [style=none] (52) at (0, 0.5) {2};
				\node [style=none] (53) at (0.5, 0.5) {3};
				\node [style=none] (54) at (-0.5, -0.5) {4};
				\node [style=none] (55) at (0, -0.5) {5};
				\node [style=none] (56) at (0.5, -0.5) {6};
			\end{pgfonlayer}
			\begin{pgfonlayer}{edgelayer}
				\fill [color=gray] (0.center) rectangle (1.center);
				\fill [color=cyan] (2.center) rectangle (3.center);
				\fill [color=green] (29.center) rectangle (33.center);
				\fill [color=cyan] (24.center) rectangle (31.center);
				\draw (0.center) to (19.center);
				\draw (20.center) to (25.center);
				\draw (31.center) to (27.center);
				\draw (36.center) to (32.center);
				\draw (42.center) to (37.center);
				\draw (47.center) to (1.center);
				\draw (1.center) to (19.center);
				\draw (18.center) to (43.center);
				\draw (44.center) to (17.center);
				\draw (16.center) to (45.center);
				\draw (46.center) to (15.center);
				\draw (0.center) to (47.center);
			\end{pgfonlayer}
		\end{tikzpicture}
		&
		\begin{tikzpicture}[scale=.5]
			\begin{pgfonlayer}{nodelayer}
				\node [style=none] (0) at (-5, 5) {};
				\node [style=none] (1) at (5, -5) {};
				\node [style=none] (2) at (-1, 1) {};
				\node [style=none] (3) at (1, -1) {};
				\node [style=none] (12) at (0, 0) {$\D$};
				\node [style=none] (15) at (-3, 5) {};
				\node [style=none] (16) at (-1, 5) {};
				\node [style=none] (17) at (1, 5) {};
				\node [style=none] (18) at (3, 5) {};
				\node [style=none] (19) at (5, 5) {};
				\node [style=none] (20) at (-5, 3) {};
				\node [style=none] (21) at (-3, 3) {};
				\node [style=none] (22) at (-1, 3) {};
				\node [style=none] (23) at (1, 3) {};
				\node [style=none] (24) at (3, 3) {};
				\node [style=none] (25) at (5, 3) {};
				\node [style=none] (27) at (-5, 1) {};
				\node [style=none] (28) at (-3, 1) {};
				\node [style=none] (29) at (1, 1) {};
				\node [style=none] (30) at (3, 1) {};
				\node [style=none] (31) at (5, 1) {};
				\node [style=none] (32) at (5, -1) {};
				\node [style=none] (33) at (3, -1) {};
				\node [style=none] (34) at (-1, -1) {};
				\node [style=none] (35) at (-3, -1) {};
				\node [style=none] (36) at (-5, -1) {};
				\node [style=none] (37) at (-5, -3) {};
				\node [style=none] (38) at (-3, -3) {};
				\node [style=none] (39) at (-1, -3) {};
				\node [style=none] (40) at (1, -3) {};
				\node [style=none] (41) at (3, -3) {};
				\node [style=none] (42) at (5, -3) {};
				\node [style=none] (43) at (3, -5) {};
				\node [style=none] (44) at (1, -5) {};
				\node [style=none] (45) at (-1, -5) {};
				\node [style=none] (46) at (-3, -5) {};
				\node [style=none] (47) at (-5, -5) {};
				\node [style=none] (48) at (2, 0) {};
				\node [style=none] (49) at (2, 0) {$\M$};
				\node [style=none] (50) at (-0.5, 0.5) {1};
				\node [style=none] (51) at (0, 2) {$\S$};
				\node [style=none] (52) at (0.5, 0.5) {2};
				\node [style=none] (53) at (1.5, 0.5) {3};
				\node [style=none] (54) at (2.5, -0.5) {4};
				\node [style=none] (55) at (-0.5, -0.5) {5};
				\node [style=none] (56) at (0.5, -0.5) {6};
			\end{pgfonlayer}
			\begin{pgfonlayer}{edgelayer}
				\fill [color=gray] (0.center) rectangle (1.center);
				\fill [color=cyan] (2.center) rectangle (3.center);
				\fill [color=green] (29.center) rectangle (33.center);
				\fill [color=cyan] (22.center) rectangle (29.center);
				\draw (0.center) to (19.center);
				\draw (20.center) to (25.center);
				\draw (31.center) to (27.center);
				\draw (36.center) to (32.center);
				\draw (42.center) to (37.center);
				\draw (47.center) to (1.center);
				\draw (1.center) to (19.center);
				\draw (18.center) to (43.center);
				\draw (44.center) to (17.center);
				\draw (16.center) to (45.center);
				\draw (46.center) to (15.center);
				\draw (0.center) to (47.center);
			\end{pgfonlayer}
		\end{tikzpicture}
		\\
		(a) & (b)
		\\&
		\\
		\begin{tikzpicture}[scale=.5]
			\begin{pgfonlayer}{nodelayer}
				\node [style=none] (0) at (-5, 5) {};
				\node [style=none] (1) at (5, -5) {};
				\node [style=none] (2) at (-1, 1) {};
				\node [style=none] (3) at (1, -1) {};
				\node [style=none] (12) at (0, 0) {$\D$};
				\node [style=none] (15) at (-3, 5) {};
				\node [style=none] (16) at (-1, 5) {};
				\node [style=none] (17) at (1, 5) {};
				\node [style=none] (18) at (3, 5) {};
				\node [style=none] (19) at (5, 5) {};
				\node [style=none] (20) at (-5, 3) {};
				\node [style=none] (21) at (-3, 3) {};
				\node [style=none] (22) at (-1, 3) {};
				\node [style=none] (23) at (1, 3) {};
				\node [style=none] (24) at (3, 3) {};
				\node [style=none] (25) at (5, 3) {};
				\node [style=none] (27) at (-5, 1) {};
				\node [style=none] (28) at (-3, 1) {};
				\node [style=none] (29) at (1, 1) {};
				\node [style=none] (30) at (3, 1) {};
				\node [style=none] (31) at (5, 1) {};
				\node [style=none] (32) at (5, -1) {};
				\node [style=none] (33) at (3, -1) {};
				\node [style=none] (34) at (-1, -1) {};
				\node [style=none] (35) at (-3, -1) {};
				\node [style=none] (36) at (-5, -1) {};
				\node [style=none] (37) at (-5, -3) {};
				\node [style=none] (38) at (-3, -3) {};
				\node [style=none] (39) at (-1, -3) {};
				\node [style=none] (40) at (1, -3) {};
				\node [style=none] (41) at (3, -3) {};
				\node [style=none] (42) at (5, -3) {};
				\node [style=none] (43) at (3, -5) {};
				\node [style=none] (44) at (1, -5) {};
				\node [style=none] (45) at (-1, -5) {};
				\node [style=none] (46) at (-3, -5) {};
				\node [style=none] (47) at (-5, -5) {};
				\node [style=none] (49) at (2, 2) {$\M$};
				\node [style=none] (50) at (-0.5, 0.5) {1};
				\node [style=none] (51) at (0, 2) {$\S$};
				\node [style=none] (52) at (0.5, 0.5) {2};
				\node [style=none] (53) at (1.5, 2.5) {3};
				\node [style=none] (54) at (2.5, 1.5) {4};
				\node [style=none] (55) at (-0.5, -0.5) {5};
				\node [style=none] (56) at (0.5, -0.5) {6};
			\end{pgfonlayer}
			\begin{pgfonlayer}{edgelayer}
				\fill [color=gray] (0.center) rectangle (1.center);
				\fill [color=cyan] (2.center) rectangle (3.center);
				\fill [color=cyan] (22.center) rectangle (29.center);
				\fill [color=green] (23.center) rectangle (30.center);
				\draw (0.center) to (19.center);
				\draw (20.center) to (25.center);
				\draw (31.center) to (27.center);
				\draw (36.center) to (32.center);
				\draw (42.center) to (37.center);
				\draw (47.center) to (1.center);
				\draw (1.center) to (19.center);
				\draw (18.center) to (43.center);
				\draw (44.center) to (17.center);
				\draw (16.center) to (45.center);
				\draw (46.center) to (15.center);
				\draw (0.center) to (47.center);
			\end{pgfonlayer}
		\end{tikzpicture}
		&
		\begin{tikzpicture}[scale=.5]
			\begin{pgfonlayer}{nodelayer}
				\node [style=none] (0) at (-5, 5) {};
				\node [style=none] (1) at (5, -5) {};
				\node [style=none] (2) at (-1, 1) {};
				\node [style=none] (3) at (1, -1) {};
				\node [style=none] (12) at (0, 0) {$\D$};
				\node [style=none] (15) at (-3, 5) {};
				\node [style=none] (16) at (-1, 5) {};
				\node [style=none] (17) at (1, 5) {};
				\node [style=none] (18) at (3, 5) {};
				\node [style=none] (19) at (5, 5) {};
				\node [style=none] (20) at (-5, 3) {};
				\node [style=none] (21) at (-3, 3) {};
				\node [style=none] (22) at (-1, 3) {};
				\node [style=none] (23) at (1, 3) {};
				\node [style=none] (24) at (3, 3) {};
				\node [style=none] (25) at (5, 3) {};
				\node [style=none] (27) at (-5, 1) {};
				\node [style=none] (28) at (-3, 1) {};
				\node [style=none] (29) at (1, 1) {};
				\node [style=none] (30) at (3, 1) {};
				\node [style=none] (31) at (5, 1) {};
				\node [style=none] (32) at (5, -1) {};
				\node [style=none] (33) at (3, -1) {};
				\node [style=none] (34) at (-1, -1) {};
				\node [style=none] (35) at (-3, -1) {};
				\node [style=none] (36) at (-5, -1) {};
				\node [style=none] (37) at (-5, -3) {};
				\node [style=none] (38) at (-3, -3) {};
				\node [style=none] (39) at (-1, -3) {};
				\node [style=none] (40) at (1, -3) {};
				\node [style=none] (41) at (3, -3) {};
				\node [style=none] (42) at (5, -3) {};
				\node [style=none] (43) at (3, -5) {};
				\node [style=none] (44) at (1, -5) {};
				\node [style=none] (45) at (-1, -5) {};
				\node [style=none] (46) at (-3, -5) {};
				\node [style=none] (47) at (-5, -5) {};
				\node [style=none] (49) at (2, 2) {$\M$};
				\node [style=none] (50) at (-0.5, 2.5) {1};
				\node [style=none] (51) at (0, 2) {$\S$};
				\node [style=none] (52) at (0, 2.5) {2};
				\node [style=none] (53) at (0.5, 2.5) {3};
				\node [style=none] (54) at (-0.5, 1.5) {4};
				\node [style=none] (55) at (0, 1.5) {5};
				\node [style=none] (56) at (0.5, 1.5) {6};
			\end{pgfonlayer}
			\begin{pgfonlayer}{edgelayer}
				\fill [color=gray] (0.center) rectangle (1.center);
				\fill [color=cyan] (2.center) rectangle (3.center);
				\fill [color=green] (23.center) rectangle (30.center);
				\fill [color=cyan] (22.center) rectangle (29.center);
				\draw (0.center) to (19.center);
				\draw (20.center) to (25.center);
				\draw (31.center) to (27.center);
				\draw (36.center) to (32.center);
				\draw (42.center) to (37.center);
				\draw (47.center) to (1.center);
				\draw (1.center) to (19.center);
				\draw (18.center) to (43.center);
				\draw (44.center) to (17.center);
				\draw (16.center) to (45.center);
				\draw (46.center) to (15.center);
				\draw (0.center) to (47.center);
			\end{pgfonlayer}
		\end{tikzpicture}
		\\
		(c) & (d)	
	\end{tabular}
	\caption{Illustration de la partie optimale, avec (a) la configuration initiale.}
	\label{fig_t1}
\end{figure}
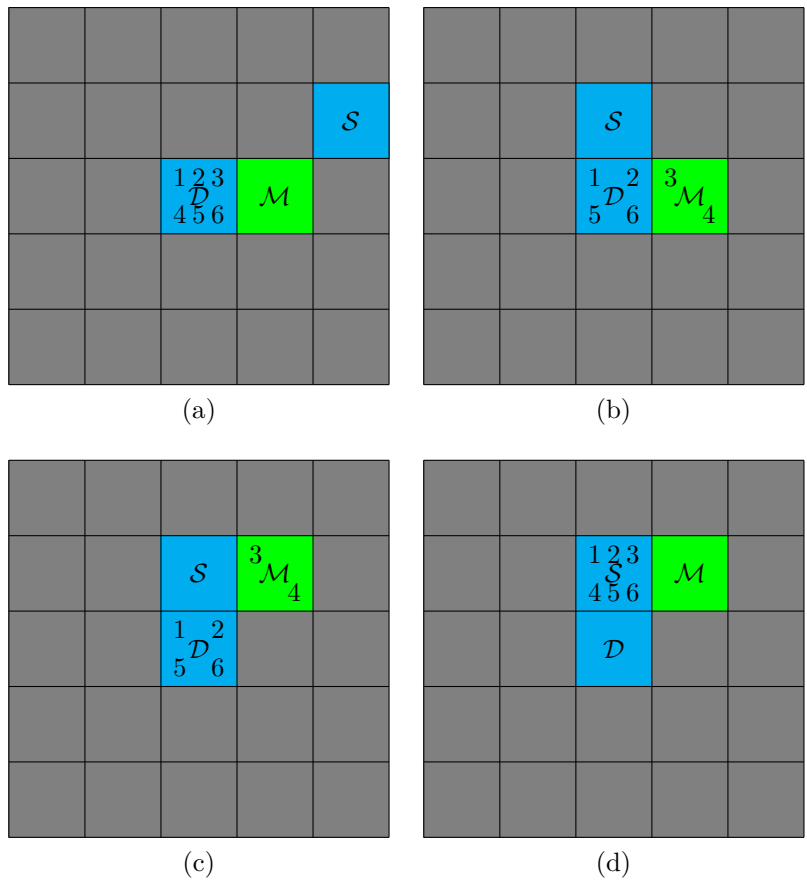
\end{dem}

\section{Probabilités de succès}\label{sec_proba}

Une fois établi que ces parties sont possibles, il est naturel de se demander à quel point elles sont probables, étant données qu’elles reposent sur des conditions initiales très spécifiques. Pour effectuer une comparaison équitable avec l’ouverture téméraire, nous considérerons l’ouverture véloce à six personnages. Au niveau de la composition du complexe, l’on avait remplacé une tuile vide par une de contrôle afin de faire fonctionner l’ouverture téméraire; l’on peut faire de même côté véloce sans aucune conséquence, en agissant sur une rangée inutilisée pour ignorer l’effet de la salle des machines.\\

Par ailleurs, comme expliqué plus tôt, toute action programmée doit être exécutée quelqu’en soient les conséquences, sans compter que l’ouverture téméraire n’a pas le luxe d’utiliser des actions $R$ pour voir où elle va. Nous étudierons donc pour chaque ouverture la probabilité de perdre immédiatement la partie en tuant un personnage dès le premier tour.

\subsection{Ouverture véloce}

Rappelons que l’ouverture véloce à six joueurs est:
\[\mathbf{1:} 1RD \sep 2RD \sep 3RD \sep 4RD \sep 5DC \sep 6DR\]
L’on remarque la présence de quatre actions regarder avant le premier déplacement, ce qui permet de voir les quatre tuiles adjacentes à $\D$ et ainsi toujours se diriger vers un $\X$ valide lorsqu’il en existe un.

\begin{prop}
	La probabilité d’être V-chanceux est de $\frac{17}{230} \simeq 7,39\%$.
\end{prop}

\begin{dem}
	Comme dit plus haut les actions $R$ permettent de trouver un $\X$ valide lorsqu’il en existe un. Comme le complexe contient $5$ salle vides (ou des machines), $2$ salles noires et $2$ tunnels, cela fait $9$ tuiles valides sur les $23$ non-bleues. La distribution étant équiprobable, la probabilité de ne pas trouver de bon $\X$ est:
	\[\frac{14}{23}\cdot\frac{13}{22}\cdot\frac{12}{21}\cdot\frac{11}{20} = \frac{13}{115}\]
	
	Une fois le $\X$ choisi, il ne reste plus qu’à viser une tuile pouvant être de sortie, et il y a une chance sur $12$ que ce soit effectivement $\S$. Soit une probabilité finale de:
	\[\left(1-\frac{13}{115}\right)\cdot\frac{1}{12} = \frac{17}{230}\]
\end{dem}

Du côté des risques, l’on ne peut perdre un personnage au premier tour que si les quatre tuiles adjacentes à $\D$ sont rouges, ce qui rend d’ailleurs la partie très difficile quelque soit la stratégie.

\begin{prop}
	La probabilité de perdre au premier tour en appliquant l’ouverture véloce est au plus $\frac{18}{1265} \simeq 1,42\%$.
\end{prop}

\begin{dem}
	Pour simplifier et puisque l’on cherche uniquement une borne supérieure, supposons que les salles rouges sont toujours fatales. Elles sont au nombre de $9$ donc la probabilité que $\D$ ne soit entouré que de rouge est de:
	\[\frac{9}{23}\cdot\frac{8}{22}\cdot\frac{7}{21}\cdot\frac{6}{20} = \frac{18}{1265}\]
\end{dem}

\begin{nb}
	L’on pourrait également se soucier de perdre au deuxième tour à cause de cette ouverture. Cela peut uniquement se produire si l’on ne peut choisir pour $\X$ qu’une salle noire, ce qui nous empêche de regarder la tuile en $[2;\!1]$ comme deuxième action du personnage 6, et que cette dernière se trouve être rouge. Si l’on craint ce problème le plus simple est de considérer que les salles noires ne sont pas viables pour $\X$, ce qui élimine le risque tout en conservant une probabilité de succès d’environ $6,62\%$.
\end{nb}

\subsection{Ouverture téméraire}

Cette fois les probabilités sont nettement moins favorables puisque, pour finir en un seul tour, l’on a besoin d’une configuration encore plus précise et d’abandonner tout recours à l’action $R$.

\begin{prop}
	La probabilité d’être T-chanceux est de $\frac{1}{276} \simeq 0,36\%$.
\end{prop}

\begin{dem}
	Puisque l’on n’a aucune information pour éclairer le choix du repère, l’on a une chance sur 23 (resp. 12) que la salle de contrôle (resp. sortie) se trouve là où l’on veut. Cela fait une probabilité finale de $\frac{1}{23}\cdot\frac{1}{12} = \frac{1}{276}$
\end{dem}

Il est évidemment bien plus probable qu’une des deux cases soit mortelle.

\begin{prop}
	La probabilité de perdre au premier tour en appliquant l’ouverture téméraire est au moins $\frac{43}{138} \simeq 31,2\%$.
\end{prop}

\begin{dem}
	Pour obtenir une borne inférieure, nous considérerons seulement les salles mortelles et piégées comme garantissant la mort d’un personnage (voir la \cref{sec_imposs}). Par construction, la tuile en position $[2;\!1]$ a une chance sur 12 d’être $\S$ puis une chance équitable d’être une des $23$ tuiles non-bleues. La probabilité qu’elle ne soit pas une des quatre tuiles interdites est donc $\frac{1}{12} + \frac{11}{12}\cdot\frac{19}{23}$. En supposant que c’est le cas, si la tuile en position $[2;\!1]$ est $\S$ alors celle en $[1;\!0]$ peut être n’importe quelle tuile non-bleue, sinon il faut retirer celle qui est en $[2;\!1]$, ce qui ne laisse que $22$ tuiles possibles. La probabilité qu’aucune des quatre tuiles interdites soit atteinte est alors:
	\[\frac{1}{12}\cdot\frac{19}{23} + \frac{11}{12}\cdot\frac{19}{23}\cdot\frac{18}{22} = \frac{95}{138}\]
	La minoration de la probabilité de défaite immédiate est donc son complémentaire, $\frac{43}{138}$.
\end{dem}

\section{Parties solo dans un contexte antagonique}\label{sec_imposs}

Après avoir considéré des parties optimalement chanceuses, l’on peut se demander s’il est toujours possible de gagner même si le hasard est optimalement défavorable.\\

Si le livret de règles invite les joueurs à choisir eux-même la composition du plateau, il déconseille toutefois d’inclure à la fois deux salles mortelles et deux vortex. La première a pour effet de tuer immédiatement tout personnage qui y rentre, et la seconde de le renvoyer immédiatement en $\D$. La raison probable de cette mise en garde est le fait que la partie devient impossible à gagner si $\D$ se trouve entouré par ces quatre tuiles. En effet les personnages ne peuvent alors pas agir sur le complexe ni réellement quitter leur position de départ parce que tout déplacement sera immédiatement soit fatal soit annulé par un vortex.\\

Si l’on s’en tient à la composition recommandée pour une partie solo en revanche, il n’est pas aussi simple de trouver une disposition initiale impossible à résoudre. Cependant, chercher une telle disposition initiale est encore un peu trop à l’avantage du joueur parce que cela suppose qu’il connaît la disposition avant de chercher à la résoudre, alors qu’au contraire il lui faudrait découvrir les cases une par une. Dans un tel cas de figure, la bonne notion de \og malchance optimale \fg{} est ce que nous appellerons un contexte antagonique. Cette notion est très proche de l’analyse compétitive des algorithmes en ligne, qui sont des algorithmes pour lesquels, comme dans notre cas, l’entrée est révélée au fur et à mesure de l’exécution \autocite{borodinOnlineComputationCompetitive1998}.

\begin{defn}
	Une partie solo dans un \textit{contexte antagonique} est un jeu à deux participants. Le premier, appelé \textit{joueur}, essaie de faire en sorte que tous les personnages s’échappent du complexe en un certain nombre de tour, exactement comme dans une partie solo standard. Le second, appelé \textit{adversaire} décide de la disposition du complexe et cherche à faire perdre le joueur, c’est-à-dire tuer au moins un personnage ou le freiner suffisamment pour qu’il ne parvienne pas à sortir assez vite.
	
	La manière dont joue l’adversaire est la suivante. Chaque fois que le joueur regarde pour la première fois le recto d’une tuile (que ce soit par l’action regarder ou bien en y faisant entrer un personnage), l’adversaire choisit de laquelle il s’agit, parmi celles qui sont cohérentes avec ses choix précédents.
\end{defn}

\begin{wrapfigure}{r}{0.4\linewidth}
	\vspace{-10mm}
	\centering
	\begin{tikzpicture}[scale=.5]
		\begin{pgfonlayer}{nodelayer}
			\node [style=none] (0) at (-5, 5) {};
			\node [style=none] (1) at (5, -5) {};
			\node [style=none] (2) at (-1, 1) {};
			\node [style=none] (3) at (1, -1) {};
			\node [style=none] (12) at (0, 0) {$\D$};
			\node [style=none] (15) at (-3, 5) {};
			\node [style=none] (16) at (-1, 5) {};
			\node [style=none] (17) at (1, 5) {};
			\node [style=none] (18) at (3, 5) {};
			\node [style=none] (19) at (5, 5) {};
			\node [style=none] (20) at (-5, 3) {};
			\node [style=none] (21) at (-3, 3) {};
			\node [style=none] (22) at (-1, 3) {};
			\node [style=none] (23) at (1, 3) {};
			\node [style=none] (24) at (3, 3) {};
			\node [style=none] (25) at (5, 3) {};
			\node [style=none] (27) at (-5, 1) {};
			\node [style=none] (28) at (-3, 1) {};
			\node [style=none] (29) at (1, 1) {};
			\node [style=none] (30) at (3, 1) {};
			\node [style=none] (31) at (5, 1) {};
			\node [style=none] (32) at (5, -1) {};
			\node [style=none] (33) at (3, -1) {};
			\node [style=none] (34) at (-1, -1) {};
			\node [style=none] (35) at (-3, -1) {};
			\node [style=none] (36) at (-5, -1) {};
			\node [style=none] (37) at (-5, -3) {};
			\node [style=none] (38) at (-3, -3) {};
			\node [style=none] (39) at (-1, -3) {};
			\node [style=none] (40) at (1, -3) {};
			\node [style=none] (41) at (3, -3) {};
			\node [style=none] (42) at (5, -3) {};
			\node [style=none] (43) at (3, -5) {};
			\node [style=none] (44) at (1, -5) {};
			\node [style=none] (45) at (-1, -5) {};
			\node [style=none] (46) at (-3, -5) {};
			\node [style=none] (47) at (-5, -5) {};
			\node [style=none] (48) at (2, 0) {};
			\node [style=none] (49) at (2, 0) {$\M$};
			\node [style=none] (50) at (0, 2) {$\O$};
			\node [style=none] (51) at (-2, 0) {$\P$};
			\node [style=none] (52) at (0, -2) {$\P$};
		\end{pgfonlayer}
		\begin{pgfonlayer}{edgelayer}
			\fill [color=gray] (0.center) rectangle (1.center);
			\fill [color=cyan] (2.center) rectangle (3.center);
			\fill [color=red] (29.center) rectangle (33.center);
			\fill [color=yellow] (22.center) rectangle (29.center);
			\fill [color=red] (28.center) rectangle (34.center);
			\fill [color=red] (34.center) rectangle (40.center);
			\fill [color=LightGrey] (27.center) rectangle (35.center);
			\fill [color=LightGrey] (35.center) rectangle (39.center);
			\fill [color=LightGrey] (39.center) rectangle (44.center);
			\fill [color=LightGrey] (21.center) rectangle (2.center);
			\fill [color=LightGrey] (3.center) rectangle (41.center);
			\draw (0.center) to (19.center);
			\draw (20.center) to (25.center);
			\draw (31.center) to (27.center);
			\draw (36.center) to (32.center);
			\draw (42.center) to (37.center);
			\draw (47.center) to (1.center);
			\draw (1.center) to (19.center);
			\draw (18.center) to (43.center);
			\draw (44.center) to (17.center);
			\draw (16.center) to (45.center);
			\draw (46.center) to (15.center);
			\draw (0.center) to (47.center);
		\end{pgfonlayer}
	\end{tikzpicture}
	\caption{Plateau de la partie antagonique après que le joueur ait forcé l’adversaire à fixer les quatre tuiles adjacentes à $\D$ en les regardant. Les cases grises indiquent les endroits où l’adversaire pourra choisir de placer la deuxième salle mortelle.}
	\label{fig_antag}
\end{wrapfigure}
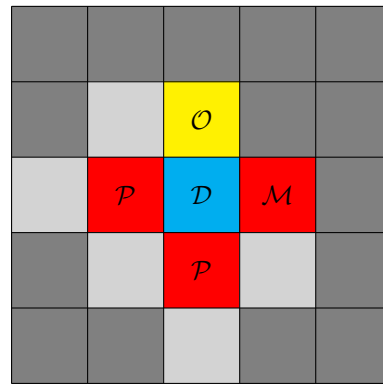

\phantom{bla}\vspace{-5mm}
\begin{prop}
	Dans un contexte antagonique, l’adversaire peut toujours forcer la défaite du joueur.
\end{prop}

\begin{dem}
	Dans un contexte antagonique, l’adversaire peut supposer que le joueur commence par regarder les quatre tuiles adjacentes au départ: s’il y a une tuile que le joueur décide de ne pas regarder, l’adversaire pourra toujours attendre qu’il le fasse pour lui montrer. Il peut alors décider que les quatre tuiles adjacentes à $\D$ soient une (des deux) salle mortelle $\M$, le vortex $\O$ et les deux salles piégées $\P$ (voir \cref{fig_antag}). Comme nous l’avons dit plus tôt, la salle mortelle tue tout personnage qui y rentre et le vortex le renvoie en $\D$, ce qui rend ces deux directions sans issue. Les salles piégées tuent tout personnage qui n’en est pas sorti à la fin de la première action qu’il entreprend depuis son entrée.\\
	
	Le seul moyen pour qu’un personnage sorte durablement de $\D$ sans qu’il n’y ait de mort est donc de se déplacer vers une salle piégée, ne plus faire d’action du tour puis, comme première action du tour suivant, se déplacer vers une des cases adjacentes (claires sur la \cref{fig_antag}). En particulier, il est obligé de se déplacer à l’aveugle car regarder est une action, ce qui fait que la salle piégée le tuerait avant qu’il n’ait le temps de se déplacer. Or l’adversaire peut alors placer la deuxième salle mortelle à l’endroit où le personnage décide d’aller, provoquant immédiatement sa mort et, subséquemment, la défaite du joueur.
\end{dem}

Une autre manière de voir les choses est de considérer un adversaire qui est obligé de fixer le plateau à l’avance. S’il fait un plateau compatible avec la \cref{fig_antag}, mais telle que la deuxième $\M$ est placée au hasard parmi les cases grises, alors le joueur a au moins 1 chance sur 5 de perdre, quelle que soit sa stratégie.

\section{Conclusion}

Sans pour autant chercher une stratégie permettant d’aborder toutes les configurations initiales possibles, nous avons vu une programmation d’ouverture permettant de gagner la partie de manière optimale (soit en deux tours) dans certains cas. Nous avons également démontré que cette ouverture menait à une défaite immédiate que dans des configurations encore plus rares que celles donnant une victoire optimale, et qui plus est difficiles à remporter quelque soit l’ouverture choisie.\\

Bien sûr, il reste à savoir si, dans l’écrasante majorité des cas où cette ouverture ne mène ni à une défaite immédiate ni à une victoire optimale, le fait de l’avoir suivie est pénalisant pour le reste de la partie. Intuitivement, l’on s’attend à ce que ce soit un peu le cas dans la mesure où l’ouverture, si elle est exécutée avec succès, envoie tous les personnages sur une même case excentrée du plateau à la fin du second tour. Or le but étant d’explorer les quatre coins pour trouver la sortie, il semble avantageux de se disperser tant que faire se peut en début de partie.

\printbibliography

\pagebreak
\appendix

\section{Preuve supplémentaire}\label{anx}

\begin{prop}
	Quelque soit le nombre de personnages, il est impossible d’obtenir une victoire partielle en coopération en un seul tour.
\end{prop}

\begin{dem}
	Comme nous l’avons dit dans le corps du texte, pour avoir une victoire partielle en un tour, il faut qu’il n’y ait qu’un seul $p\in\Pi$ qui aurait eu besoin d’un second déplacement pour atteindre $\S$, et qu’il meure avant que les autres personnages n’essaient de décaler $\S$ hors du plateau. Le tout sachant que $p$ ne peut pas changer de salle dans cet intervalle de temps puisqu’il a déjà consommé son déplacement et que, les autres étant en $\S$ ou en $\D$, personne ne peut le pousser. Or la mort d’un personnage ne peut être due qu’aux quatre salles suivantes:
	\begin{itemize}
		\item le \textit{bain d’acide}, qui tue le premier arrivé quand un second entre, ne peut pas tuer $p$ si tous les autres personnages sont ailleurs;
		\item la \textit{salle innondable} ne tue un personnage qu’à la fin du tour suivant l’entrée, donc ne peut tuer personne dans une partie à un seul tour;
		\item la \textit{salle mortelle} tue immédiatement, donc $p$ n’a pas pu aider les autres (en poussant quelqu’un ou en rendant $\S$ adjacente à $\D$ par une action $C$) depuis cette salle;
		\item la \textit{salle piégée}, qui est la seule qu’il reste à traiter.
	\end{itemize}
	
	La salle piégée tue tout personnage qui n’en est pas sorti à la fin de la première action qu’il entreprend depuis celle-ci. Etant donné que le $D$ est forcément consommé pour sortir de $\D$, tout personnage atteignant une salle piégée en mourra s’il entreprend une action. \\
	
	Si la dernière action de $p$ est un $C$, alors il ne peut y avoir d’autres membres dans $\Pi$ car $p$ ne pousse pas et que tout membre de $\Pi$ ne peut rejoindre $\S$ qu’en étant poussé par un autre. Or il faut au moins deux $C$ pour rendre $\S$ adjacente à $\D$ ce qui est alors impossible à faire.\\
	
	La dernière action doit donc être le fait de pousser le dernier autre membre de $\Pi$ dans $\S$, depuis la salle piégée donc. Ce n’est pas possible si la salle piégée est adjacente à $\D$ car $\S$ est forcément trop loin. Sinon, pour que $p$ atteigne la salle piégée, il faut qu’il se fasse pousser par un autre membre de $\Pi$, ce qui en laisse donc forcément un hors de $\S$ à la mort de $p$.
\end{dem}

%
%
%
%

\end{document}